\documentclass[numberedheadings]{aip-cp}

\usepackage[numbers,sort&compress]{natbib}
\usepackage{fancyhdr}
\newenvironment{DIFnomarkup}{}{}

\usepackage{rotating}
\usepackage{graphicx}
\usepackage{comment}

\begin{document}

\title{Post-CCSD(T) corrections to bond distances and vibrational frequencies: the power of $\Lambda$}

\author[aff1,aff2,aff4]{Maciej Spiegel}
\author[aff1,aff4]{Emmanouil Semidalas}
\author[aff1]{Jan M. L. Martin}
\eaddress{gershom@weizmann.ac.il}
\author[aff3]{Megan R. Bentley}
\author[aff3]{John F. Stanton}

\affil[aff1]{Dept. of Molecular Chemistry and Materials Science, Weizmann Institute of Science, 7610001 Re\d{h}ovot, Israel.}
\affil[aff2]{Department of Pharmacognosy and Herbal Medicines, Wroclaw Medical University, Borowska 211a, 50-556 Wroc{\l}aw, Poland.}
\affil[aff3]{Quantum Theory Project, Department of Chemistry, University of Florida, Gainesville, FL 32611, USA.}
\affil[aff4]{Equally contributing first authors}

\maketitle

\begin{abstract}
The importance of post-CCSD(T) corrections as high as CCSDTQ56 for ground-state spectroscopic constants ($D_e$, $\omega_e$, $\omega_ex_e$, and $\alpha_e$) has been surveyed for a sample of two dozen mostly heavy-atom diatomics spanning a broad range of static correlation strength. While CCSD(T) is known to be an unusually felicitous `Pauling point' between accuracy and computational cost, performance leaves something to be desired for molecules with strong static correlation. We find CCSDT(Q)$_\Lambda$ to be the next `sweet spot' up, of comparable or superior quality to the much more expensive CCSDTQ. A similar comparison applies to CCSDTQ(5)$_\Lambda$ vs. CCSDTQ5, while CCSDTQ5(6)$_\Lambda$ is essentially indistinguishable from CCSDTQ56.
A composite of CCSD(T)-X2C/ACV5Z-X2C with [CCSDT(Q)$_\Lambda$ -- CCSD(T)]/cc-pVTZ or even cc-pVDZ basis sets appears highly effective for computational vibrational spectroscopy. Unlike CCSDT(Q) which breaks down for the ozone vibrational frequencies, CCSDT(Q)$_\Lambda$ handles them gracefully.

\end{abstract}

\section{Introduction}

The CCSD(T) method\cite{Rag89,Wat93}, coupled cluster with all single and double substitutions\cite{Purvis1982} and a quasiperturbative correction\cite{Rag89,Wat93} for connected triple excitations, is also known as "the gold standard of quantum chemistry" (expression first coined by T. H. Dunning, Jr. in a 2000 lecture). It is indeed a particularly felicitous compromise between accuracy and computational cost, and for most applications CCSD(T) can reliably yield chemically accurate results. However, as became apparent during high-accuracy computational thermochemistry (e.g., the Weizmann-4\cite{jmlm200,jmlm205} and HEAT\cite{heat1,heat2,heat3,heat4} approaches) and computational spectroscopy studies, its excellent performance is the result of an error compensation between neglect of higher-order triples and connected quadruples. In the  presence of static correlation, this compensation becomes erratic; hence, for benchmark accuracy, one needs to go beyond CCSD(T).

Already in 2001, Musia{\l} et al.\cite{Musia2001} found that connected {\em quintuples} (!) could contribute quite nontrivially to stretching frequencies for triple bonds. This was further explored by Th{\o}gersen and Olsen\cite{Thogersen2004}  and by Ruden et al.\cite{Ruden2004}. 

Cortez et al.\cite{Cortez2007} considered the vibrational spectrum of water, and found: that CCSD(T) was in error by just a few cm$^{-1}$; that CCSDT (coupled cluster with all single, double, and triple substitutions) did not improve on this; and that CCSDT(Q) (i.e., CCSDT with a quasiperturbative correction for connected quadruples\cite{Bomble2005}) approached the full CI limit to better than 1~cm$^{-1}$. (See also K\'allay and Gauss\cite{KallayFreq2004}.) These observations are consistent with Ref.\cite{jmlm260} which defines the HFREQ27 vibrational frequencies benchmark, where CCSD(T) at the basis set limit was found to have an RMSD of 4.6 cm$^{-1}$. The molecules in HFREQ27 are a mix of heavy-atom and hydrogen-containing diatomics with small polyatomics, excluding cases known to exhibit strong static correlation. 

 In 2010, Karton and Martin\cite{jmlm230} investigated the application of the W4 composite scheme\cite{jmlm200} to spectroscopic constants (by finite difference of pointwise W4 energies). W4 contains terms through CCSDTQ5 (coupled cluster with all connected single through quintuple substitutions, a.k.a. CCSDTQP for `pentuple' excitations\footnote{As `pentuple' after `quadruple' switches from Latin to Greek, the corresponding author prefers `quintuple'. Both notations can however be found in the literature.}), albeit in progressively smaller basis sets as the substitution level increases. The importance of each term for frequencies was tested in that paper through truncation: the authors found that stopping at CCSD(T) caused an RMS error of just 5 cm$^{-1}$ for diatomic hydride stretching frequencies, but  11 cm$^{-1}$ for nonhydrides. CCSDT yielded an improvement to 3 cm$^{-1}$ for hydrides but a deterioration to 15 cm$^{-1}$ for nonhydrides, while truncation at CCSDTQ (i.e. only omitting the CCSDTQ5 step) brought down errors drastically to 0.6 cm$^{-1}$ for hydrides and 2.4 cm$^{-1}$ for nonhydrides. The cases where the quintuples contributions were significant all involve multiple bonds, but the converse is not necessarily true.

It was thus shown once again that CCSD(T) is a Pauling point or, if the reader likes, a `sweet spot' between accuracy and computational cost. Prospects for going much beyond that in accuracy, on even a semi-routine basis, seem fairly bleak given the daunting $n^4N^6$ computational cost scaling of CCSDTQ (where $n$ represents the number of electrons correlated and $N$ the size of the basis set).

In the context of a computational thermochemistry study reconsidering both the W4-11 thermochemistry benchmark\cite{jmlm235,jmlm273} and W4 theory itself [E. Semidalas, A. Karton, and J. M. L. Martin, to be published] so-called lambda coupled cluster methods\cite{lambdastanton1,lambdastanton2,lambdabartlett1,lambdabartlett2}
seemed to show promise (see also Refs.\cite{Thorpe2023,Franke2023}). However, we concluded that the n-particle space convergence of properties other than equilibrium thermochemistry, such as spectroscopic constants, might shed more light on the problem.

We present such a study here, and shall show that there is another `sweet spot' beyond CCSD(T), namely CCSDT(Q)$_\Lambda$. In addition, we shall demonstrate that CCSDTQ(5)$_\Lambda$ can essentially be considered full CI quality for most purposes.

\begin{table}[h!]
\label{tab:norm}
\caption{Diagnostics at the CCSDTQ/cc-pVDZ level}
\centering

\centering
\begin{tabular}{lrrrrrrrrrr}\hline
                   & \multicolumn{1}{l}{${\cal T}_1$} & \multicolumn{1}{l}{${\cal T}_2$} & \multicolumn{1}{l}{${\cal T}_3$} & \multicolumn{1}{l}{${\cal T}_4$}  & \multicolumn{1}{l}{max$T_1$} & \multicolumn{1}{l}{max$T_2$} & \multicolumn{1}{l}{max$T_3$} & \multicolumn{1}{l}{max$T_4$} & \multicolumn{1}{l}{$D_1$} & \multicolumn{1}{l}{$D_2$}  \\\hline
CN                 & \textbf{0.099} & 0.100 & \textbf{0.020} & 0.005 & \textbf{0.263} & 0.119 & 0.018 & 0.003 & \textbf{0.154} & 0.195 \\
Cl$_2$                & 0.007 & 0.070 & 0.009 & 0.001 & 0.017 & 0.121 & 0.003 & 0.000 & 0.021 & 0.184 \\
BF                 & 0.016 & 0.075 & 0.010 & 0.002 & 0.030 & 0.119 & 0.006 & 0.002 & 0.037 & 0.228 \\
SO                 & 0.016 & 0.073 & 0.012 & 0.002 & 0.035 & 0.110 & 0.012 & 0.001 & 0.044 & 0.156 \\
C$_2$                 & 0.037 & \textbf{0.154} & \textbf{0.030} & \textbf{0.008} & 0.085 & \textbf{0.399} & 0.019 & 0.006 & 0.087 & \textbf{0.385} \\
CO                 & 0.017 & 0.077 & 0.014 & 0.003 & 0.030 & 0.086 & 0.013 & 0.004 & 0.044 & 0.160 \\
CS                 & 0.024 & 0.093 & 0.019 & 0.005 & 0.037 & 0.115 & 0.016 & 0.005 & 0.056 & 0.183 \\
BH                 & 0.013 & 0.109 & 0.011 & 0.002 & 0.016 & 0.122 & 0.005 & 0.001 & 0.026 & 0.266 \\
S$_2$                 & 0.014 & 0.081 & 0.014 & 0.003 & 0.033 & 0.120 & 0.007 & 0.002 & 0.017 & 0.158 \\
CH                 & 0.010 & 0.083 & 0.009 & 0.001 & 0.012 & 0.108 & 0.006 & 0.001 & 0.016 & 0.203 \\
PN                 & 0.018 & 0.098 & 0.019 & 0.005 & 0.021 & 0.148 & 0.010 & 0.003 & 0.037 & 0.211 \\
O$_2$                 & 0.013 & 0.068 & 0.009 & 0.002 & 0.033 & 0.123 & 0.005 & 0.003 & 0.013 & 0.133 \\
OH                 & 0.005 & 0.056 & 0.004 & 0.001 & 0.007 & 0.053 & 0.002 & 0.000 & 0.012 & 0.129 \\
P$_2$                 & 0.016 & 0.105 & 0.021 & 0.005 & 0.030 & 0.155 & 0.009 & 0.004 & 0.033 & 0.218 \\
HF                 & 0.004 & 0.051 & 0.003 & 0.001 & 0.006 & 0.053 & 0.002 & 0.000 & 0.014 & 0.115 \\
B$_2$                 & 0.036 & \textbf{0.154} & \textbf{0.037} & \textbf{0.010} & 0.072 & \textbf{0.355} & \textbf{0.027} & \textbf{0.016} & 0.073 & \textbf{0.395} \\
N$_2$                 & 0.011 & 0.080 & 0.012 & 0.003 & 0.018 & 0.113 & 0.006 & 0.004 & 0.026 & 0.176 \\
BN                 & \textbf{0.082} & \textbf{0.143} & \textbf{0.039} & \textbf{0.012} & \textbf{0.215} & \textbf{0.346} & \textbf{0.046} & \textbf{0.012} & \textbf{0.196} & \textbf{0.328} \\
F$_2$                 & 0.009 & 0.065 & 0.008 & 0.001 & 0.022 & 0.209 & 0.004 & 0.001 & 0.032 & 0.245 \\
NO                 & 0.030 & 0.074 & 0.013 & 0.002 & 0.094 & 0.138 & 0.022 & 0.001 & 0.054 & 0.186 \\
NH                 & 0.008 & 0.061 & 0.005 & 0.001 & 0.014 & 0.048 & 0.002 & 0.000 & 0.010 & 0.142 \\
SiO & 0.029 & 0.088 & {\bf 0.020} & 0.006 & 0.058 & 0.088 & 0.022  & {\bf 0.011} & 0.066 & 0.195\\
\hline
O$_3$ & 0.030 & 0.089 & \textbf{0.020} & 0.005 & 0.086 & \textbf{0.314} & \textbf{0.023} & 0.002 & 0.078 & 0.278 \\\hline
\end{tabular}

\tablenote{Reference geometries were CCSDTQ5(6)$_\Lambda$/cc-pVDZ. Conspicuously large values in a given column are bolded.}

\end{table}

In addition, as a polyatomic `proof of principle', we will consider the vexing problem of the ozone vibrational frequencies. 
Stanton, Magers et al.\cite{Stanton1989,Magers1989} found in 1989 that the asymmetric stretch in particular is inordinately sensitive to the treatment of triple excitations. This is caused primarily by a low-lying $[...](4b_2)^2(6a_1)^2(2b_1)^2$ configuration, with redistribution within the $\pi$ space a close second. Refs.\cite{Stanton1989,Magers1989}  considered quasiperturbative approaches such as CCSD[T] as well as approximate iterative approaches such as CCSDT-1 and CCSDT-2\cite{Noga1987}. Watts et al.\cite{Watts1991} upgraded this to full CCSDT, and found that the asymmetric stretch differed by 100 cm$^{-1}$ from CCSDT-2 and by 41 cm$^{-1}$ from the more advanced CCSDT-3; in contrast, CCSD(T) (first reported in the original paper\cite{Rag89} defining this method) lay 100 cm$^{-1}$ below CCSDT, while the more approximate CCSD[T] yielded an unphysical symmetry breaking —  making this method worse than omitting triples altogether (in this high-static-correlation regime).

Watts et al.\cite{Watts1991} also suggested that connected quadruples, $T_4$, might have a significant impact on the final result. This was first considered by Kucharski and Bartlett\cite{Kucharski1999} using the reduced-cost, factorized CCSDT(Q$_f$) approach. They found that the effect of $T_4$ on the total was highly dependent on the quality of the underlying approximation to CCSDT: on the asymmetric stretch $\omega_3$, this varies from +43 cm$^{-1}$ for CCSD(TQ$_f$) via --6 cm$^{-1}$ for CCSDT-3(Q$_f$) to --16 cm$^{-1}$ for CCSDT(Q$_f$). 

Pabst et al.\cite{Pabst2010} found that the popular CC2 approximation\cite{Christiansen1995,Haettig2003,Koehn2003} to CCSD  predicts a barrier-less dissociation of ozone to separate atoms -- a plain chemical absurdity, which illustrates the challenging character of this system from an electronic structure point of view.

Karton and Martin in their 2010 paper\cite{jmlm230} carried out pointwise W4 calculations and found harmonic frequencies $\omega_1$=1133.9, $\omega_2$=715.2, and $\omega_3$=1067.1 cm$^{-1}$, compared to values extracted from experiment\cite{Barbe2002} (see also Ref.\cite{Tyuterev1999}) of 1133.3(4), 715.0(4), and 1087.3(3) cm$^{-1}$, respectively. Upon truncating the W4 model at CCSDT and CCSDTQ levels and comparing the two, it appeared that connected quadruples affect the stretching frequencies by about 15 cm$^{-1}$.

In terms of multireference calculations, Peterson and coworkers\cite{Xie2000} were able to achieve agreement to 13 cm$^{-1}$ with CAS(12,9)-icMRCI+Q/cc-pVQZ, where CAS(12,9) refers to CASSCF\cite{Roos1980} with an active space of 12 electrons in 9 orbitals (i.e., full valence minus the 2s orbitals on each oxygen), icMRCI stands for internally contracted multireference configuration interaction\cite{Werner1988,Knowles1988}, and Q refers to adding the multireference analog\cite{Siegbahn1983} of the Davidson correction\cite{Davidson1974}. This remaining error was removed in a subsequent study by Szalay and coworkers\cite{Holka2010,Tyuterev2013} where the active space was enlarged to CAS(18,12) full valence. While this may seem to be the answer to a quantum chemist's prayer, full-valence CAS-icMRCI is no longer a feasible approach for molecules with more than four nonhydrogen atoms, owing to the factorial cost scaling of CASSCF with the numbers of electrons and active orbitals.

Li and Paldus\cite{Li1999} were able to obtain an at least qualitatively correct answer using 3R-CCSD (three-reference CCSD), where the three reference determinants were [...](HOMO)$^2$(LUMO)$^0$, [...](HOMO)$^0$(LUMO)$^2$, and [...](HOMO)$^1$(LUMO)$^1$; if they omitted the third reference, poor results were obtained. In a related vein, Szalay and Bartlett\cite{Szalay1993} found in the original paper defining the AQCC (averaged quadratic coupled cluster) method that 2R-CI (two-reference CI) and 2R-ACPF (averaged coupled pair functional\cite{Gdanitz1988}) failed, but 2R-AQCC held its own.
Hanauer and K\"ohn, in their paper\cite{Hanauer2012} on internally contracted multireference coupled cluster, found that CAS(2,2)-icCCSDT with adequately large basis sets yielded good agreement with experiment.

\section{Computational methods}

The diatomic molecule electronic structure calculations were carried out using the arbitrary-order coupled cluster implementation\cite{mrcc8,mrcc63,mrcc65,mrcc69} in the MRCC program system of K\'allay and coworkers\cite{MRCC}. 

Total energies were converged to 10$^{-11}$ hartree or better; coupled cluster jobs were run in `sequential restart' fashion where, e.g., CCSDT takes initial $T_1$ and $T_2$ amplitudes from the converged CCSD calculation, CCSDTQ in turn uses the converged CCSDT amplitudes as initial guesses for the $T_1$, $T_2$, and $T_3$ amplitudes, and so forth. For the open-shell species, unrestricted Hartree-Fock references were used throughout except where explicitly stated otherwise.

Basis sets used were the Dunning correlation consistent\cite{Dun89} cc-pVDZ, cc-pVTZ, and cc-pVQZ family.

Total energies for each diatomic were evaluated at 21 points at a spacing of 0.01 \AA, ten points each forward and backward around the  experimental bond distance rounded to 2 decimal places. The latter was taken from Huber and Herzberg\cite{Hub79}.

Potential curves were then fitted, and a Dunham analysis\cite{Dunham1932} carried out, using the program \texttt{dunham} written by one of us (JMLM) in his graduate student days. First the lowest energy point $r_i$ among the input points was found; then polynomials of increasing order in $r-r_i$ were fitted, and the goodness of fit subjected to a standard F-test. The highest order polynomial for which the addition of an extra term was still $\geq$99.9\% statistically significant was retained (in practice, this was order 7--9 for the molecules considered here). Then the minimum $r_e$ of this polynomial was found by the Newton-Raphson approach, the polynomial re-expanded in the dimensionless Dunham variable $(r-r_e)/r_e$, and finally the Dunham analysis proper carried out, including propagation of the fit uncertainties.

\begin{table}[h]
\caption{Root mean square deviations (RMSD), mean signed deviations (MSD), and mean absolute deviations (MAD) for diatomic spectroscopic constants from the CCSDTQ5(6)$_\Lambda$ level with the cc-pVDZ basis set}\label{tab:VDZ}

{\footnotesize
\begin{tabular}{lrrrrrrrrrrr}
\hline
             & \multicolumn{2}{c}{$r_e$ (\AA)}                      & \multicolumn{2}{c}{$\omega_e$ (cm$^{-1}$)}         & \multicolumn{2}{c}{$\omega_ex_e$ (cm$^{-1}$)}      & \multicolumn{2}{c}{$\alpha_e$ (cm$^{-1}$)}         & \multicolumn{3}{c}{$D_e$ (kcal/mol))}                       \\
             & \multicolumn{1}{l}{RMSD} & \multicolumn{1}{l}{MSD} & \multicolumn{1}{l}{RMSD} & \multicolumn{1}{l}{MSD} & \multicolumn{1}{l}{RMSD} & \multicolumn{1}{l}{MSD} & \multicolumn{1}{l}{RMSD} & \multicolumn{1}{l}{MSD} & \multicolumn{1}{l}{RMSD} & \multicolumn{1}{l}{MSD} & MAD    \\ 
\cline{1-12}
CCSD                & 0.01215 & -0.00975 & 59.6 & 49.7 & 1.34 & -1.08 & 0.00494 & -0.00268 & 7.763                & -6.193               & 6.193 \\
CCSD[T]             & 0.00814 & -0.00213 & 36.3 & 8.1  & 5.33 & 1.08  & 0.00162 & -0.00011 & 1.704                & 0.235                & 0.811 \\
CCSD(T)             & 0.00395 & -0.00281 & 20.2 & 14.2 & 1.89 & 0.09  & 0.00147 & -0.00059 & 0.955                & -0.719               & 0.741 \\
CCSD(T)$_\Lambda$   & 0.00395 & -0.00287 & 32.4 & 19.2 & 0.57 & 0.30  & 0.00171 & -0.00078 & 1.289                & -0.960               & 0.960 \\
CCSDT               & 0.00240 & -0.00184 & 14.3 & 10.4 & 0.42 & -0.20 & 0.00046 & -0.00028 & 0.931                & -0.737               & 0.737 \\
CCSDT[Q]            & 0.00222 & -0.00045 & 10.1 & 4.5  & 0.53 & -0.22 & 0.00031 & -0.00016 & 0.670                & -0.386               & 0.388 \\
CCSDT(Q)            & 0.00078 & -0.00003 & 6.5  & -1.7 & 0.45 & 0.12  & 0.00040 & 0.00007  & 0.248                & 0.104                & 0.125 \\
ditto w/o BN        &  0.00039 & 0.00012 & 3.4 & -0.5 & 0.16 & 0.03 & 0.00014 & 0.00000    & 0.181                & 0.071                & 0.092\\
CCSDT(Q)$_A$        & 0.00160 & -0.00004 & 7.4  & 1.7  & 0.53 & -0.20 & 0.00019 & -0.00008 & 0.366                & -0.104               & 0.156 \\
CCSDT(Q)$_B$        & 0.00101 & 0.00005  & 8.0  & 0.0  & 0.55 & -0.17 & 0.00027 & -0.00010 & 0.199                & -0.029               & 0.098 \\
ditto w/o BN        & 0.00044 & -0.00015 & 4.0 & 1.5 & 0.24 & -0.07 & 0.00018 & -0.00006   & 0.127                & 0.004                & 0.068 \\
\hline 
CCSDT(Q)$_\Lambda$  & 0.00031 & -0.00008 & 3.0  & 0.4  & 0.22 & -0.07 & 0.00017 & -0.00005 & 0.106                & 0.004                & 0.057 \\
ditto w/o BN        & 0.00027 & -0.00012 & 2.2  & 0.9  & 0.22 & -0.08 & 0.00015 & -0.00007 & 0.100                & -0.005               & 0.051 \\
\hline
CCSDTQ              & 0.00041 & -0.00026 & 2.5  & 1.5  & 0.11 & -0.02 & 0.00007 & -0.00002 & 0.128                & -0.080               & 0.083 \\
ditto w/o BN        & 0.00039 & -0.00024 & 2.5 & 1.4 & 0.08 & -0.04 & 0.00006   & -0.00003 & 0.124                & -0.075               & 0.078\\
CCSDTQ[5]           & 0.00037 & -0.00004 & 2.3  & 0.7  & 0.13 & -0.05 & 0.00006 & -0.00002 & 0.075                & -0.026               & 0.035 \\
CCSDTQ(5)           & 0.00037 & -0.00004 & 2.3  & 0.7  & 0.13 & -0.05 & 0.00006 & -0.00002 & 0.075                & -0.026               & 0.035 \\
CCSDTQ(5)$_A$       & 0.00032 & -0.00007 & 2.0  & 0.9  & 0.12 & -0.05 & 0.00006 & -0.00003 & 0.075 & -0.033 & 0.036      \\
CCSDTQ(5)$_B$       & 0.00037 & -0.00005 & 2.2  & 0.7  & 0.14 & -0.06 & 0.00006 & -0.00003 & 0.082 & -0.032 & 0.037      \\
CCSDTQ(5)$_\Lambda$ & 0.00007 & -0.00003 & 0.5  & 0.2  & 0.02 & -0.01 & 0.00002 & -0.00001 & 0.016                & -0.006               & 0.011 \\
CCSDTQ5             & 0.00005 & -0.00003 & 0.3  & 0.2  & 0.01 & 0.00  & 0.00001 & 0.00000  & 0.018                & -0.009               & 0.010 \\
CCSDTQ5[6]            & 0.00003 & -0.00002 & 0.2  & 0.1  & 0.01 & 0.00  & 0.00000 & 0.00000  & 0.014 & -0.007 & 0.008  \\
CCSDTQ5(6)            & 0.00005 & -0.00001 & 0.4  & 0.1  & 0.03 & -0.01 & 0.00001 & 0.00000  & 0.006 & -0.002 & 0.003  \\
CCSDTQ5(6)$_A$        & 0.00004 & -0.00001 & 0.3  & 0.1  & 0.02 & -0.01 & 0.00001 & 0.00000  & 0.007 & -0.003 & 0.003  \\
CCSDTQ5(6)$_B$        & 0.00004 & -0.00001 & 0.4  & 0.1  & 0.03 & -0.01 & 0.00001 & 0.00000  & 0.007 & -0.003 & 0.003  \\
CCSDTQ5(6)$_\Lambda$  & \multicolumn{1}{l}{REF}  & \multicolumn{1}{l}{REF} & \multicolumn{1}{l}{REF}  & \multicolumn{1}{l}{REF} & \multicolumn{1}{l}{REF}  & \multicolumn{1}{l}{REF} & \multicolumn{1}{l}{REF}  & \multicolumn{1}{l}{REF} & 0.002                    & 0.000                   & 0.001                     \\
CCSDTQ56               & \multicolumn{1}{l}{—}    & \multicolumn{1}{l}{—}   & \multicolumn{1}{l}{—}    & \multicolumn{1}{l}{—}   & \multicolumn{1}{l}{—}    & \multicolumn{1}{l}{—}   & \multicolumn{1}{l}{—}    & \multicolumn{1}{l}{—}   & \multicolumn{1}{l}{REF}  & \multicolumn{1}{l}{REF} &  \multicolumn{1}{l}{REF}  \\
\hline
\end{tabular}
}

\end{table}

The diatomic molecules included were a mix of:
    (a) closed-shell 1st-row: C$_2$, CO, N$_2$, BN,  BF, F$_2$, BH, CH, NH, OH, HF;
      (b) closed-shell heavy atom 2nd-row and mixed: PN, P$_2$,  SiO,  CS, HCl;
    (c) open-shell: B$_2$, O$_2$, S$_2$, SO, CN.
These molecules span a broad range of nondynamical correlation character, from dominated by dynamical correlation (such as HF and BF) to pathologically strong static correlation in C$_2$ and BN, and all gradations in between. This is illustrated by the vector norms and maximum elements of the $T_1$, $T_2$, $T_3$, and $T_4$ vectors of the molecules (Table \ref{tab:norm}). The ${\cal T}_1$ diagnostic of Lee and Taylor\cite{Lee1989} (see also Ref.\cite{LeeT1revisited}) as well as the $D_1$ diagnostic\cite{JanssenD1} of Nielsen and Janssen and the $D_2$ diagnostic of the same authors\cite{JanssenD2}. By extension from Ref.\cite{Lee1989} we also obtained 
\begin{equation}
    {\cal T}_n = ||T_n||/N_e^{1/2}
\end{equation}
where $T_n$ is the cluster amplitudes vector for $n$-tuple substitutions and $N_e$ represents the number of (in this case, valence) electrons being correlated. For closed-shell cases and $n=1$, Eq.(1) is equivalent to the original ${\cal T}_1$ diagnostic if core electrons are excluded.

The non-$\Lambda$ levels of coupled cluster theory considered include CCSD\cite{Purvis1982}, CCSD[T] (a.k.a., CCSD+T(CCSD)),\cite{Noga1986} CCSD(T),\cite{Rag89,Wat93} CCSDT,\cite{Noga1986} CCSDT[Q],\cite{Kucharski1998} CCSDT(Q),\cite{mrcc8} CCSDT(Q)$_A$,\cite{mrcc65} CCSDT(Q)$_B$,\cite{mrcc65} CCSDTQ,\cite{Kucharski1992}
CCSDTQ(5),\cite{mrcc65} and CCSDTQ5.\cite{Musia2002}

Lambda approaches were introduced independently from different considerations by Stanton et al.\cite{lambdastanton1,lambdastanton2} and by Kucharski and Bartlett\cite{lambdabartlett1,lambdabartlett2}.

In order to be able to consider the broadest variety of approximate methods, UHF references were used throughout. Initially we had included the NO molecule in our sample as well: however, it quickly became apparent that its calculated potential curves were too noisy to allow for Dunham fits of acceptably high order. This is a consequence of the UHF solution for this molecule bifurcating\cite{Szalay2004} near the equilibrium bond distance. The corresponding ROHF-reference potential curves were devoid of similar noise and could be fitted to 8$^{th}$ order without problems. As this, however, would mean skipping NO for part of the data columns where no ROHF implementation is available, we deleted NO from the analysis.

Single-reference equilibrium geometries and harmonic vibrational frequencies of ozone were determined using the CFOUR program package\cite{Cfour2020}. The correlation consistent cc-pVDZ basis set of Dunning and coworkers\cite{Dun89} was used with a restricted Hartree-Fock (RHF) reference for calculations ranging from CCSD\cite{Purvis1982} to CCSDTQ5\cite{Musia2002}. Aside from approximate coupled cluster approaches such as asymmetric (lambda-based) methods like CCSD(T)$_{\Lambda}$, more well-known symmetric counterparts (like CCSD(T)\cite{Rag89,Wat93} and its precursor CCSD+T(CCSD),\cite{Noga1986} a.k.a. CCSD[T]) were included in this study. Calculations requiring up to full quadruple excitations, CCSDTQ\cite{Kucharski1992}, were carried out with the NCC program developed by Matthews and coworkers\cite{Cfour2020}, while those incorporating higher excitations used MRCC\cite{MRCC}. 

Total energies were converged to at least the ninth decimal place for CCSDTQ(5), CCSDTQ(5)$_\Lambda$, and CCSDTQ5, and beyond the tenth decimal places for all lower levels. For levels of theory possessing implemented analytic gradients (up to CCSDTQ, excluding CCSDT[Q]\cite{Kucharski1998} and CCSDT(Q)$_{\Lambda}$), harmonic force fields were calculated via numerical differentiation of analytically evaluated forces. Otherwise, harmonic frequencies were determined through double numerical differentation of energies. Comparisons between harmonic force fields evaluated at the CCSD(T) level indicated that these partially numerical methods yield frequencies within 0.1 cm$^{-1}$ of the `exact' values obtained through analytical second derivatives\cite{Gauss1997,Harding2008} of CCSD(T). 

Multireference calculations were performed using MOLPRO 2022.3\cite{Molpro2020}, which was also employed for the large basis CCSD(T) steps in some of the composite schemes discussed below.

\section{Results and discussion}

\subsection{Diatomics\label{sec:diatomics}}

\begin{table}[h!]
\caption{Root mean square deviations (RMSD), mean signed deviations (MSD), and mean absolute deviations (MAD) for diatomic spectroscopic constants from the CCSDTQ(5)$_\Lambda$ level with the cc-pVTZ basis set}
\label{tab:VTZ}
\tabcolsep7pt
{\footnotesize
\begin{tabular}{lrrrrrrrrrrr}
\hline
                    & \multicolumn{2}{c}{$r_e$ (\AA)} & \multicolumn{2}{c}{$\omega_e$ (cm$^{-1}$)} & \multicolumn{2}{c}{$\omega_ex_e$ (cm$^{-1}$)} & \multicolumn{2}{c}{$\alpha_e$ (cm$^{-1}$)} & \multicolumn{3}{c}{$D_e$ (kcal/mol))}\\
                    & RMSD                  & MSD                    & RMSD                 & MSD                 & RMSD                  & MSD                   & RMSD                & MSD                  &
                    \multicolumn{1}{l}{RMSD} & \multicolumn{1}{l}{MSD} & MAD    \\ 
                    \hline
CCSD                      & 0.0114 & -0.0093 & 60.0 & 50.3 & 1.26 & -0.93 & 0.00487 & -0.00273 & 8.814 & -7.220 & 7.220 \\
CCSD{[}T{]}               & 0.0068 & -0.0019 & 32.6 & 8.1 & 4.21 & 0.74 & 0.00122 & -0.00020 &  1.597 & 0.442 & 0.644  \\
{\bf CCSD(T) }            & 0.0034 & -0.0024 & 20.1 & 13.4 & 1.37 & -0.01 & 0.00131 & -0.00066 & 0.714 & -0.460 & 0.494 \\
CCSD(T)$_\Lambda$         & 0.0035 & -0.0026 & 31.0 & 18.8 & 1.52 &  0.00    & 0.00170 & -0.00099 & 1.189 & -0.842 & 0.842 \\
CCSDT                     & 0.0024 & -0.0018 & 14.0 & 10.5 & 0.41 & -0.21 & 0.00057 & -0.00034 & 1.071 & -0.829 & 0.829 \\
CCSDT{[}Q{]}              & 0.0020 & -0.0006 & 9.3 & 4.6 & 0.47 & -0.21 & 0.00029 & -0.00020 & 0.881 & -0.598 & 0.599 \\
CCSDT(Q)                  & 0.0009 & 0.0000 & 5.2 & -1.9 & 0.53 & 0.13 & 0.00039 & 0.00008 & 0.265 & 0.111 & 0.124 \\
ditto w/o BN & 0.0003 & 0.0001 & 2.0 & -0.9 &  0.14 & 0.02 & 0.00012 & 0.00001 & 0.187 & 0.075 & 0.088\\
CCSDT(Q)$_A$              & 0.0014 & 0.0000 & 5.9 & 1.2 & 0.37 & -0.14 & 0.00013 & -0.00006 & 0.368 & -0.118 & 0.158 \\
CCSDT(Q)$_B$              & 0.0008 & 0.0000 & 7.6 & -0.6 & 0.50 & -0.15 & 0.00021 & -0.00008 & 0.203 & -0.039 & 0.096 \\
ditto w/o BN & 0.0003 & -0.0001 & 2.8 & 0.9 & 0.21 & -0.06 & 0.00015 & -0.00005 & 0.123 & -0.006 & 0.065 \\
\hline
{\bf CCSDT(Q)$_\Lambda$}  & 0.0002 & -0.0001 & 2.4 & 0.3 & 0.19 & -0.07 & 0.00015 & -0.00004 & 0.090 & -0.006 & 0.045 \\
ditto w/o BN & 0.0002 & -0.0001 & 1.4 & 0.7 & 0.19 & -0.07 & 0.00013 & -0.00006 & 0.091 & -0.010 & 0.044 \\
\hline
CCSDTQ                    & 0.0004 & -0.0002 & 2.1 & 1.4 & 0.08 & -0.02 & 0.00006 & -0.00002 & 0.122 & -0.081 & 0.082 \\
CCSDTQ{[}5{]}             & 0.0004 & 0.0000 & 2.1 & 0.5 & 0.12 & -0.03 & 0.00005 & -0.00002 & 0.100 & -0.044 & 0.049 \\
CCSDTQ(5)                 & 0.0004 & 0.0000 & 2.1 & 0.5 & 0.12 & -0.03 & 0.00005 & -0.00002 & 0.100 & -0.044 & 0.049 \\
CCSDTQ(5)$_A$             & 0.0003 & 0.0000 & 1.9 & 0.6 & 0.11 & -0.04 & 0.00005 & -0.00002 & 0.102 & -0.050 & 0.050 \\
CCSDTQ(5)$_B$             & 0.0004 & 0.0000 & 2.0 & 0.5 & 0.13 & -0.04 & 0.00005 & -0.00002 & 0.109 & -0.049 & 0.052 \\
CCSDTQ(5)$_\Lambda$       &  REF     &  REF      &  REF    &  REF   &  REF   &  REF    &  REF      &  REF       & REF & REF & REF \\\hline
\end{tabular}
}

\end{table}

For the small cc-pVDZ basis set, we can easily go all the way up to connected sextuple excitations. 
Table \ref{tab:VDZ} presents the performance statistics in the cc-pVDZ basis set of the various coupled cluster approximations for the following four quantities: bond distance $r_e$, harmonic frequency $\omega_e$, first anharmonicity constant $\omega_ex_e$, and rotation-vibration coupling constant $\alpha_e$. (Note that these quantities require at most fourth derivatives, for $\omega_ex_e$, and that we have enough data points that we can reliably fit sixth- or higher-order expansions.)

First of all, the difference between CCSDTQ5(6)$_\Lambda$ and fully iterative CCSDTQ56 is negligible, at 0.002 kcal/mol RMSD. Only for one diatomic, C$_2$, does it reach 0.01 kcal/mol. We thus deem it justified to use CCSDTQ5(6)$_\Lambda$ as our quasi-full CI reference. The complete sextuples contribution CCSDTQ56 - CCSDTQ5 works out to 0.018 kcal/mol RMS and -0.009 kcal/mol MSD. The large RMSD/MAD ratio indicates an outlier (for an unbiased normal distribution, RMSD/MAD$=\sqrt{\pi/2}\approx$5/4 \cite{Geary35,jmlm235}), which turns out to be C$_2$ at 0.068 kcal/mol.

\begin{table}[h!]
\caption{Comparison for selected diatomics between cc-pVQZ and cc-pVTZ basis sets for differences (\AA,cm$^{-1}$) with CCSDTQ(5)$_\Lambda$ }
\label{tab:VQZ}

{\footnotesize
\begin{tabular}{l|rrrr|rrrr}
\hline & \multicolumn{4}{c}{cc-pVQZ} & \multicolumn{4}{c}{cc-pVTZ} \\
B$_2$ & $\Delta R_e$ & $\Delta\omega_e$ & $\Delta\omega_ex_e$ & $\Delta\alpha_e$ & $\Delta R_e$ & $\Delta\omega_e$ & $\Delta\omega_ex_e$ & $\Delta\alpha_e$ \\
\hline
CCSD(T) & -0.00408 & 8.20 & -0.004 & 0.00002 & -0.00424 & 8.49 & -0.016 & 0.00001 \\
CCSD(T)$_\Lambda$ & -0.00196 & 5.76 & -0.010 & -0.00001 & -0.00221 & 5.95 & -0.022 & -0.00002 \\
CCSDT & -0.00059 & -0.52 & 0.050 & 0.00010 & -0.00061 & -0.42 & 0.041 & 0.00009 \\
CCSDT(Q) & -0.00036 & -0.01 & 0.021 & 0.00004 & -0.00032 & -0.05 & 0.018 & 0.00004 \\
CCSDT(Q)/B & -0.00067 & 0.95 & 0.034 & 0.00003 & -0.00062 & 0.87 & 0.032 & 0.00003 \\
CCSDT(Q)$_\Lambda$ & -0.00043 & 0.48 & 0.024 & 0.00003 & -0.00043 & 0.47 & 0.021 & 0.00003 \\
CCSDTQ & -0.00002 & -0.31 & 0.012 & 0.00002 & -0.00001 & -0.33 & 0.010 & 0.00002 \\
CCSDTQ(5) & -0.00013 & 0.47 & -0.003 & -0.00001 & -0.00012 & 0.45 & -0.003 & -0.00001 \\
CCSDTQ(5)$_B$ & -0.00013 & 0.44 & -0.001 & -0.00001 & -0.00012 & 0.42 & -0.002 & -0.00001 \\
\hline
C$_2$ & $\Delta R_e$ & $\Delta\omega_e$ & $\Delta\omega_ex_e$ & $\Delta\alpha_e$ & $\Delta R_e$ & $\Delta\omega_e$ & $\Delta\omega_ex_e$ & $\Delta\alpha_e$ \\
\hline
CCSD(T) & -0.00195 & 15.44 & -1.030 & -0.00061 & -0.00204 & 15.63 & -0.996 & -0.00059 \\
CCSD(T)$_\Lambda$ & -0.00077 & 17.77 & -1.299 & -0.00085 & -0.00092 & 18.12 & -1.271 & -0.00084 \\
CCSDT & -0.00218 & 17.53 & -0.886 & -0.00058 & -0.00217 & 17.50 & -0.870 & -0.00057 \\
CCSDT(Q) & 0.00031 & -1.67 & -0.098 & -0.00004 & 0.00032 & -1.55 & -0.100 & -0.00004 \\
CCSDT(Q)/B & -0.00016 & 1.93 & -0.086 & -0.00007 & -0.00015 & 2.02 & -0.083 & -0.00008 \\
CCSDT(Q)$_\Lambda$ & 0.00006 & 2.65 & -0.482 & -0.00028 & 0.00006 & 2.75 & -0.479 & -0.00028 \\
CCSDTQ & -0.00049 & 2.99 & -0.052 & -0.00004 & -0.00048 & 2.93 & -0.046 & -0.00004 \\
CCSDTQ(5) & 0.00010 & -0.60 & 0.051 & 0.00002 & 0.00010 & -0.63 & 0.052 & 0.00002 \\
CCSDTQ(5)$_B$  & 0.00005 & -0.27 & 0.051 & 0.00002 & 0.00005 & -0.30 & 0.051 & 0.00002 \\
\hline
N$_2$ & $\Delta R_e$ & $\Delta\omega_e$ & $\Delta\omega_ex_e$ & $\Delta\alpha_e$ & $\Delta R_e$ & $\Delta\omega_e$ & $\Delta\omega_ex_e$ & $\Delta\alpha_e$ \\
\hline
CCSD(T) & -0.00095 & 12.58 & -0.319 & -0.00024 & -0.00098 & 12.89 & -0.335 & -0.00025 \\
CCSD(T)$_\Lambda$ & -0.00124 & 15.86 & -0.374 & -0.00029 & -0.00125 & 15.94 & -0.391 & -0.00030 \\
CCSDT & -0.00163 & 22.50 & -0.577 & -0.00045 & -0.00162 & 22.55 & -0.593 & -0.00046 \\
CCSDT(Q) & 0.00015 & -2.73 & 0.157 & 0.00009 & 0.00014 & -2.65 & 0.160 & 0.00009 \\
CCSDT(Q)/B & 0.00010 & -2.00 & 0.140 & 0.00008 & 0.00009 & -1.96 & 0.146 & 0.00008 \\
CCSDT(Q)$_\Lambda$ & 0.00005 & -0.98 & 0.070 & 0.00004 & 0.00004 & -0.88 & 0.070 & 0.00004 \\
CCSDTQ & -0.00024 & 3.82 & -0.141 & -0.00010 & -0.00024 & 3.75 & -0.142 & -0.00010 \\
CCSDTQ(5) & 0.00002 & -0.66 & 0.060 & 0.00003 & 0.00002 & -0.68 & 0.061 & 0.00003 \\
CCSDTQ(5)$_B$  & 0.00001 & -0.47 & 0.049 & 0.00003 & 0.00002 & -0.49 & 0.049 & 0.00003 \\
\hline
P$_2$& $\Delta R_e$ & $\Delta\omega_e$ & $\Delta\omega_ex_e$ & $\Delta\alpha_e$ & $\Delta R_e$ & $\Delta\omega_e$ & $\Delta\omega_ex_e$ & $\Delta\alpha_e$ \\
\hline
CCSD(T) & -0.00324 & 9.40 & -0.160 & -0.00005 & -0.00360 & 10.45 & -0.186 & -0.00006 \\
CCSD(T)$_\Lambda$ & -0.00527 & 14.60 & -0.213 & -0.00007 & -0.00562 & 15.58 & -0.241 & -0.00008 \\
CCSDT & -0.00512 & 14.85 & -0.188 & -0.00007 & -0.00528 & 15.25 & -0.194 & -0.00007 \\
CCSDT(Q) & 0.00023 & -1.22 & 0.060 & 0.00001 & 0.00021 & -1.18 & 0.069 & 0.00001 \\
CCSDT(Q)/B & 0.00005 & -0.73 & 0.057 & 0.00001 & 0.00003 & -0.73 & 0.066 & 0.00001 \\
CCSDT(Q)$_\Lambda$  & -0.00040 & 1.01 & 0.006 & 0.00000 & -0.00045 & 1.10 & 0.010 & 0.00000 \\
CCSDTQ & -0.00084 & 2.76 & -0.045 & -0.00002 & -0.00086 & 2.77 & -0.044 & -0.00002 \\
CCSDTQ(5) & -0.00005 & -0.23 & 0.024 & 0.00001 & -0.00002 & -0.32 & 0.025 & 0.00001 \\
CCSDTQ(5)$_B$ & -0.00008 & -0.11 & 0.020 & 0.00000 & -0.00005 & -0.19 & 0.021 & 0.00000 \\
\hline
BN & $\Delta R_e$ & $\Delta\omega_e$ & $\Delta\omega_ex_e$ & $\Delta\alpha_e$ & $\Delta R_e$ & $\Delta\omega_e$ & $\Delta\omega_ex_e$ & $\Delta\alpha_e$ \\
\hline

CCSD(T) & -0.01004 & 44.54 & 5.456 & 0.00081 & -0.01110 & 42.10 & 6.313 & 0.00130\\
CCSD(T)$_\Lambda$ &-0.00376 & 77.56 & 4.088 & -0.00127& -0.00486 & 82.44 & 6.511 & -0.00073\\
CCSDT & -0.00263 & 7.77 & 0.991	& 0.00059& -0.00256 & 6.91 & 0.937 & 0.00060\\
CCSDT(Q) & -0.00443 & -20.58 & 2.561 & 0.00175& -0.00409 & -23.25 & 2.444 & 0.00178\\
CCSDT(Q)/B & 0.00323 & -35.31 & -2.098 &-0.00056& 0.00352 & -33.99 & -2.167 & -0.00071\\
CCSDT(Q)$_\Lambda$ & 0.00069 & -8.98 & 0.071 & 0.00034& 0.00074 & -9.39 & -0.007 & 0.00033\\
CCSDTQ & -0.00078 & 2.51 & 0.294 & 0.00017& -0.00076 & 2.27 & 0.275 & 0.00017\\
CCSDTQ(5) & 0.00132 & -1.02 & -0.181 & -0.00008& 0.00126 & -0.83 & -0.173 & -0.00009\\
CCSDTQ(5)$_B$ & 0.00134 & -2.51 & -0.368 & -0.00010& 0.00129 & -2.22 & -0.370 & -0.00012\\
\hline
\end{tabular}
}
\tablenote{Due to resource constraints, post-CCSDTQ for BN was obtained by the pointwise additivity approximation E[CCSDTQ(5)$_\Lambda$/cc-pVQZ]$\approx$} 
\tablenote{E[CCSDTQ/cc-pVQZ]+E[CCSDTQ(5)$_\Lambda$/cc-pVQZ(no g)]-E[CCSDTQ/cc-pVQZ(no g)]}
\tablenote{~~~}
\tablenote{~~~}
\tablenote{~~~}
\tablenote{~~~}
\end{table}

Second, it is immediately apparent from the spectroscopic constants that the effect of sextuples is negligible for all but the most demanding applications — 0.00005 \AA\ for $r_e$, 0.3 cm$^{-1}$ for harmonic frequencies, 0.01 cm$^{-1}$ for the anharmonic constant, and 0.00001 cm$^{-1}$  for the rovibrational coupling constant.

\begin{table}[h!]
\caption{RMSD and MSD (Root Mean Square and Mean Signed Deviations in \AA,cm$^{-1}$) of spectroscopic constants from experiment with various composite schemes}
\label{tab:vsExpt}

{\footnotesize
\begin{tabular}{ll|rrr|rrr}

\hline
& & RMSD & RMSD & RMSD & MSD & MSD & MSD\\
\hline
CCSD(T) component & post-CCSD(T) correction & $\Delta r_e$ & $\Delta \omega_e$ & $\Delta \omega_ex_e$ & $\Delta r_e$ & $\Delta \omega_e$ & $\Delta \omega_ex_e$\\
\hline
CCSD(T)/ACV5Z (NR, no CV)    & nil & 0.00306 & 6.73 & 0.422 & 0.00246 & 1.97 & -0.299 \\
CCSD(T)/ACV5Z (CV+NR) & nil & 0.00136 & 10.40 & 0.378 & -0.00090 & 8.90 & -0.232 \\
CCSD(T)-X2C/ACV5Z-X2C (CV+R) & nil & 0.00131 & 9.09 & 0.380 & -0.00088 & 7.47 & -0.238  \\
CCSD(T)-X2C/ACV6Z-X2C (CV+R) & nil & 0.00163 & 9.63 & 0.517 & -0.00147 & 9.15 & -0.347 \\
CCSD(T)-X2C/ACV\{5,6\}Z-X2C (CV+R) & nil & 0.00233 & 11.03 & 0.748 & -0.00219 & 10.71 & -0.406 \\
CCSD(T)-X2C/ACV5Z-X2C (CV+R) & CCSDT/cc-pVTZ & 0.00148  & 12.54  &  0.392 & -0.00091 & 8.95 & -0.227 \\
CCSD(T)-X2C/ACV5Z-X2C (CV+R) & CCSDT(Q)/cc-pVDZ$^a$ & 0.00120  & 3.70 & 0.202 & 0.00091 & -2.28 & 0.033  \\
CCSD(T)-X2C/ACV5Z-X2C (CV+R) & CCSDT(Q)$_\Lambda$/cc-pVDZ$^a$ & 0.00100  & 2.46 & 0.235 & 0.00073 & -1.00 & -0.033  \\
CCSD(T)-X2C/ACV5Z-X2C (CV+R) & CCSDT(Q)/cc-pVTZ$^a$ & 0.00128  & 3.64   & 0.177 & 0.00093 & -2.42 & 0.063  \\
CCSD(T)-X2C/ACV5Z-X2C (CV+R) & CCSDT(Q)$_\Lambda$/cc-pVTZ$^a$ & 0.00103  & 2.09 & 0.198 & 0.00072 & -0.87 & -0.018 \\
CCSD(T)-X2C/ACV5Z-X2C (CV+R) & CCSDT(Q)$_\Lambda$/cc-pVTZ & 0.00102  & 2.09 & 0.189 & 0.00073 & -0.98 & -0.013 \\
CCSD(T)-X2C/ACV5Z-X2C (CV+R) & CCSDTQ(5)$_\Lambda$/cc-pVTZ & 0.00112  & 2.07 & 0.167 & 0.00082 & -1.41 & 0.031 \\
CCSD(T)/ACV\{5,6\}Z+$\Delta$X2C/ACV5Z (CV+R)$^b$ & ditto & 0.00058  & 1.93 & 0.200 & -0.00043 & 1.04 & -0.103 \\

CCSD(T)-X2C/ACV5Z-X2C (CV+R) & CCSDT(Q)$_\Lambda$/cc-pVQZ & 0.00103 & 1.92 & 0.183 & 0.00074 & -0.97 & -0.018 \\
CCSD(T)-X2C/ACV\{5,6\}Z-X2C (CV+R) & ditto & 0.00077 & 2.75 & 0.384 & -0.00049 & 1.54 & -0.128\\
CCSD(T)/ACV\{5,6\}Z+$\Delta$X2C/ACV5Z (CV+R)$^b$ & ditto & 0.00071 & 2.49 & 0.257 & -0.00054 & 1.78 & -0.143\\

CCSD(T)-X2C/ACV5Z-X2C (CV+R) & CCSDT(Q)$_\Lambda$/cc-pVQZ +  \\
&[CCSDTQ(5)$_\Lambda$ - \\
&CCSDT(Q)$_\Lambda$]/cc-pVTZ & 0.00114 & 1.94 & 0.127 & 0.00083 & -1.41 & 0.011 \\
CCSD(T)-X2C/ACV\{5,6\}Z-X2C (CV+R) & ditto & 0.00061 & 2.16 & 0.422 & -0.00041 & 0.88 & -0.063\\
CCSD(T)/ACV\{5,6\}Z+$\Delta$X2C/ACV5Z (CV+R)$^b$ & ditto & 0.00056 & 1.92 & 0.275 & -0.00045 & 1.08 & -0.075\\
\hline
\hline
\end{tabular}
}
\tablenote{NR: Non-Relativistic, R: Relativistic, CV: core-valence correlation}
\tablenote{post-CCSD(T) corrections: CCSDT/BASIS refers to E[CCSDT/BASIS] -- E[CCSD(T)/BASIS]}
\tablenote{All expt. reference data from the compilation of Irikura\cite{Irikura2007}, except B$_2$, where they are taken from Ref.\cite{Hub79}}
\tablenote{(a) Excluding CN, NO, and SO, as UHF-CCSDT(Q) unsuitable for these species and ROHF-CCSDT(Q) not implemented}
\tablenote{(b) The $\Delta$X2C/ACV5Z term indicates an additive relativistic correction [CCSD(T)-X2C/ACV5Z-X2C (CV+R) - CCSD(T)/ACV5Z (CV)], where X2C is the `exact two-component' approach\cite{X2C}.}

\end{table}

Third, the difference between CCSDTQ(5)$_\Lambda$ and fully iterative CCSDTQ5 is similarly small — and the two tiny differences  even partially compensate, as reflected in the CCSDTQ5(6)$_\Lambda$ -- CCSDTQ(5)$_\Lambda$ difference. In light of the fact that 
CCSDTQ(5)$_\Lambda$ is vastly cheaper than fully iterative CCSDTQ5 — asymptotically scaling as $O(n^5N^6)$ rather than $O(n^5N^7)$ — this would seem to be a good choice for a reference level.
The superiority of CCSDTQ(5)$_\Lambda$ over CCSDTQ(5) was previously noted for the 8-valence electron diatomics\cite{jmlm253}.

Fourth, the apparent effect of connected quintuple excitations is to lengthen bonds by an average of 0.0002 \AA, concomitantly with an average decrease in the harmonic frequency by 1.3 cm$^{-1}$ and a negligible average increase in the anharmonicity of just 0.02 cm$^{-1}$. However, for some molecules like N$_2$ and PN, the difference may reach a less trivial 4-5 cm$^{-1}$.

Fifth, the other approximate quintuples treatments such as CCSDTQ(5) are all clearly inferior in quality to CCSDTQ(5)$_\Lambda$, and it is not even clear that they are superior to CCSDTQ.

Sixth, the performance difference between CCSDTQ and CCSDT(Q)$_\Lambda$ is remarkably small, and much of that difference can in fact be attributed to the pathologically multireference BN diatomic --- which is a taxing test for any 
electron correlation method, and \textit{a fortiori} for any quasiperturbative correction. Except for the cases of CN and BN, it is not obvious that CCSDT(Q)$_\Lambda$ agrees less well with the reference than does CCSDTQ: indeed, upon removal of BN, statistics for 
harmonic frequencies at the CCSDT(Q)$_\Lambda$ level are superior to those of CCSDTQ. Again, the reduced scaling by $O(n^4N^5)$ instead of $O(n^4N^6)$ is a great boon.

Seventh, CCSDT(Q)$_\Lambda$ is clearly superior to CCSDT(Q). CCSDT(Q)$_B$ represents no obvious improvement over CCSDT(Q). Superior performance of CCSDT(Q)$_\Lambda$ over other quasiperturbative quadruples options was first noted for a smaller sample of eight-valence-electron species in Ref.\cite{jmlm253}.

For further comparison, let us turn to the cc-pVTZ basis set (Table \ref{tab:VTZ}). The observations above are largely seen here as well. 
CCSDT(Q)$_\Lambda$ is again found to be a felicitous compromise between performance and computational cost: the shortcomings of the other approximate connected quadruples methods are actually clearest in the anharmonicity constant, and to a lesser extent in the rovibrational coupling constant and the bond distance. The large RMSD/MSD ratio for the $D_e$ contribution indicates an outlier, which is B$_2$ at 0.36 kcal/mol. Note from Table \ref{tab:norm} that B$_2$ has an unusually large $T_4$ amplitude of 0.016, with only the pathologically multireference BN molecule even coming close at 0.012.

In terms of contributions to the dissociation energy, predictably there are differences between the basis sets for CCSD, CCSD(T), and to a lesser extent CCSDT. However, from CCSDT(Q) onward, the differences with the reference are remarkably similar between the cc-pVDZ and cc-pVTZ basis sets, which reflects the observation from Refs.\cite{jmlm200,jmlm205} that basis set convergence of terms in the cluster expansion becomes ever faster as one walks up the substitution ladder.

Comparison between CCSDT and CCSDTQ reveals that connected quadruples lengthen bonds across the board, by an average of 0.0016 \AA, and reduce harmonic frequencies by an average of 8.9 cm$^{-1}$. With the exception of BN, anharmonicities are slightly increased as well (0.18 cm$^{-1}$ on average) by connected quadruples. 

Full CCSDT minus CCSD(T) is more of a mixed bag, with bond contractions seen in P$_2$, PN, N$_2$, and O$_2$, and lengthening in the other diatomics (translating into an average lengthening by 0.0018 \AA). Significantly, the error statistics for harmonic frequencies are not that dissimilar between CCSDT and CCSD(T): it is in the anharmonicity and rovibrational coupling constants that CCSDT is clearly superior. The other approximate triples contributions are inferior to CCSD(T) in performance.

The RMSD of 14 cm$^{-1}$ for CCSD(T) seems to be at odds with the 5 cm$^{-1}$ reported in Ref.\cite{jmlm260} for HFREQ27. However, said set was long on species such as C$_2$H$_4$, N$_2$H$_4$, CH$_3$OH, and the like, where the X-H stretching frequencies are much easier to describe without high-order connected excitations.

Is there a need to consider larger basis sets? Comparing Tables \ref{tab:VDZ} and \ref{tab:VTZ}, we note that the difference in statistics becomes quite small for CCSDT and insignificant for higher levels. 
This would seem to make proceeding further to cc-pVQZ redundant; nevertheless, by way of verification, 
we did carry out  CCSDTQ(5)$_\Lambda$/cc-pVQZ calculations
for B$_2$, C$_2$, N$_2$, and P$_2$. The differences from CCSDTQ(5)$_\Lambda$ for those four molecules are compared between cc-pVTZ and cc-pVQZ in Table \ref{tab:VQZ}.
Their trends clearly parallel what has been reported above --- most saliently, that agreement of CCSDT(Q)$_\Lambda$ with CCSDTQ(5)$_\Lambda$ meets or exceeds that of fully iterative CCSDTQ.

\begin{DIFnomarkup}
\begin{table}[h!]
\centering
\caption{Convergence of  geometry (\AA, degrees) and harmonic frequencies (cm$^{-1}$) of ozone along the coupled cluster series}
\label{tab:ozone}

\begin{tabular}{lrrrrr}
\hline
                    & $\omega_1$ & $\omega_2$ & $\omega_3$ & $r_{OO}$ & $\theta$ \\
                    &$a_1$ & $a_1$ & $b_2$ &\\
                    \hline
                    \multicolumn{6}{c}{cc-pVDZ}\\
                    \hline
                    CCSDTQ(5)$_\Lambda$
                    & 1072.6          & 687.8           & 990.8           & 1.29268       & 116.217         \\

                   & $\Delta\omega_1$ & $\Delta\omega_2$ & $\Delta\omega_3$ & $\Delta r_{OO}$ & $\Delta \theta$ \\

\hline
CCSD                & 178.1           & 65.4            & 246.0           & -0.03370      & 1.070           \\
CCSD[T]             & 13.1            & -1.9            & -789.5          & -0.00256      & 0.570           \\
CCSD(T)             & 44.9            & 16.4            & -14.1           & -0.00828      & 0.424           \\
CCSDT-1a & -19.8 & -1.1 & -303.1 & 0.00063 & 0.127 \\
CCSDT-1b &\ 9.7 & 6.0 & 88.9 & -0.00213 & 0.202 \\
CCSDT-2 & 70.8 & 25.0 & 178.7 & -0.01133 & 0.394 \\
CCSDT-3 & 62.7 & 19.4 & 114.1 & -0.01013& 0.490 \\
CCSD(T)$_\Lambda$   & 72.4            & 24.8            & 69.8            & -0.01235      & 0.571           \\
CCSDT               & 53.9            & 17.7            & 74.1            & -0.00927      & 0.272           \\
CCSDT[Q]            & 19.4            & 8.0             & 50.2           & -0.00340      & 0.069           \\
CCSDT(Q)            & -15.6           & -5.6            & -85.0           & 0.00217       & -0.016          \\
CCSDTQ-3 & 20.8 & 7.6 & 42.2 & -0.00324 & 0.049 \\
CCSDT(Q)$_\Lambda$  & 6.3             & 1.6             & 12.2            & -0.00072      & 0.040           \\
CCSDTQ              & 10.6            & 3.3             & 16.3            & -0.00165      & 0.039           \\
CCSDTQ[5]           & -3.4            & -0.5            & -12.2           & 0.00035       & -0.019          \\
CCSDTQ(5)           & -3.4            & -0.5            & -12.2           & 0.00035       & -0.019          \\
\textbf{CCSDTQ5} & \textbf{+0.3} & \textbf{+0.1} & \textbf{+1.6} & \textbf{-0.00003} & \textbf{+0.004} \\
\hline
CAS(18,12)-icAQCC & -1.1 & -0.5 & 0.3 & 0.00077 & -0.024 \\
CAS(12, 9)-icAQCC & 1.6 & 1.8 & 0.5 & -0.00073 & 0.016 \\
CAS(18,12)-icACPF & -2.1 & -1.4 & -1.7 & 0.00118 & -0.032\\
CAS(18,12)-icMRCI & 3.4 & 3.9 & 9.2 & -0.00144 & 0.024\\
CAS(18,12)-icMRCI+Q & 0.2 & 0.3 & -0.0 & 0.00008 & -0.004\\
CAS(2,2)-MRCCSDT\cite{Hanauer2012} & 9.4 & 2.2 & 14.2 & -0.00128 & 0.043\\
                    \hline
                    \multicolumn{6}{c}{cc-pVTZ}\\
                    \hline
CAS(18,12)-icAQCC & 1111.1 & 700.8 & 1058.9 & 1.28339 & 116.541 \\
                   & $\Delta\omega_1$ & $\Delta\omega_2$ & $\Delta\omega_3$ & $\Delta r_{OO}$ & $\Delta \theta$ \\
\hline
CCSD(T) & 42.0 & 14.9 & -4.6 & -0.00789 & 0.407\\
CCSDT & 52.1 & 16.7 & 58.2 & -0.00940 & 0.280\\
CCSDT[Q] & 19.5 & 7.9 & 51.2 & -0.00409 & 0.080 \\
CCSDT(Q) & -17.0 & -5.8 & -87.2 & 0.00195 & -0.037\\
CCSDT(Q)$_\Lambda$ & 3.3 & 0.6 & 7.3 & -0.00070 & 0.017\\
                    \hline
                    \multicolumn{6}{c}{cc-pVQZ}\\
                    \hline
CAS(18,12)-icAQCC & 1129.4 & 710.5 & 1083.2 & 1.27626 & 116.728 \\
                   & $\Delta\omega_1$ & $\Delta\omega_2$ & $\Delta\omega_3$ & $\Delta r_{OO}$ & $\Delta \theta$ \\
\hline
CCSD(T) & 40.4 & 14.1 & -2.0 & -0.00745 & 0.390\\
CCSDT &  51.0 & 16.0 & 53.3 & -0.00903 & 0.274 \\
CCSDT[Q] & 18.5 & 7.5 & 49.6 & -0.00392 & 0.076\\
                    \hline
                    \multicolumn{6}{c}{cc-pV5Z}\\
                    \hline
CAS(18,12)-icAQCC & 1133.9 & 713.0 & 1089.1 & 1.27460 & 116.760 \\
Experiment\cite{Barbe2002,Tyuterev1999} & 1133.3(5) & 714.9(4) & 1087.2(3) & 1.27276(15) & 116.754(3)\\
\hline
\end{tabular}
\end{table}
\end{DIFnomarkup}

\subsection{Composite schemes and comparison with experiment}
 
Some might argue that thus far, we have been comparing to putative full CI limits for finite basis sets, and that we ought to consider comparison with experiment instead. For all diatomics considered here except BN, reliable experimental spectroscopic constants are available from Irikura\cite{Irikura2007} or from the older, but still widely used, Huber and Herzberg reference book\cite{Hub79}. To this end, we obtained potential curves using a simple composite scheme in which all-electron CCSD(T) including  X2C (exact two-component\cite{X2C}) corrections  was supplemented with an additive post-CCSD(T) correction of the form, e.g., E[CCSD(T)/LARGE+CV+REL]+E[CCSDT(Q)/SMALL] - E[CCSD(T)/SMALL]. The resulting error statistics can be found in Table~\ref{tab:vsExpt}.

We use here the X2C recontractions, provided in the MOLPRO basis set library, of the aug-cc-pCV5Z basis set\cite{Peterson2002} combined with aug-cc-pV5Z\cite{Kendall1992} on hydrogen.

Without post-CCSD(T) corrections, bond distances are systematically too short (and hence, due to Badger's Rule\cite{Badger1934}) harmonic frequencies systematically blue-shifted. RMS error in $\omega_e$ reaches nearly 10 cm$^{-1}$, and 0.5 cm$^{-1}$ in the first anharmonic correction. Expanding the basis set further in fact increases these errors. Including just a CCSDT correction somewhat reduces bond compression, but does not improve statistics overall. 

A CCSDT(Q)/cc-pVTZ correction slightly overcorrects bond lengths, hence leading to slightly too low harmonic frequencies, but a much improved RMSD of 3.6 cm$^{-1}$, while that for anharmonicity constants improves to 0.18 cm$^{-1}$.
 Substituting CCSDT(Q)$_\Lambda$, however, reduces the systematic underestimate of harmonic frequencies to less than 1 cm$^{-1}$, and hence RMSD further to 2.1 cm$^{-1}$. This appears to be the `Pareto optimum' for this simple combination: upgrading the post-CCSD(T) correction further to CCSDTQ(5)$_\Lambda$ does not result in any significant further improvement in the harmonic frequencies, though the RMSD in the anharmonicity constants is reduced slightly further, for 0.19 to 0.17 cm$^{-1}$. These `diminishing returns' do not justify the massive cost increase.

Increasing the underlying basis set for CCSD(T) to 6Z causes major near-linear dependence issues for some molecules when X2C is enabled; for the remainder, it appears that with the CCSDTQ(5)$_\Lambda$ correction, RMSD for bond distances can be reduced to 0.0006 \AA\ (mean signed error -0.0004 \AA). 

We attempted both increasing the basis set for (Q)$_\Lambda$ to cc-pVQZ and additionally including a CCSDTQ(5)$_\Lambda$--CCSDT(Q)$_\Lambda$ correction with the cc-pVTZ basis set. Neither is able to do substantially better than 2 cm$^{-1}$ for harmonic frequencies --- which 
appears to be the practical limit in this "CCSD(T) plus corrections" framework. Additionally improving the CCSD(T) part by considering a ACV\{5,6\}Z basis set extrapolation does not improve the frequencies much, although bond lengths are now substantially better than 0.001 \AA\ RMS.

This relative insensitivity to the quality of the basis set in the post-CCSD(T) step prompts the question whether one might be able to substitute the small cc-pVDZ basis set and still obtain useful results. Indeed, a composite of CCSD(T)-X2C/ACV5Z-X2C with CCSDT(Q)$_\Lambda$/cc-pVDZ yields surprisingly good RMSD=2.5 cm$^{-1}$ on harmonic frequencies and 0.001 \AA\ on bond distances. This holds out a tantalizing possibility for polyatomic molecules large enough that CCSDT(Q)$_\Lambda$/cc-pVTZ would be cost-prohibitive.

 \subsection{The vibrational frequencies of ozone}

\begin{figure}[h!]
\label{fig:megan}
\caption{Harmonic vibrational frequencies of ozone at various levels of coupled cluster theory using the cc-pVDZ basis set, followed by a pictoral representation of the dependence of $\omega_{3}$. The CASSCF(18,12)-AQCC/cc-pVDZ result is shown schematically by the red line. 
}
\includegraphics[width=5in]{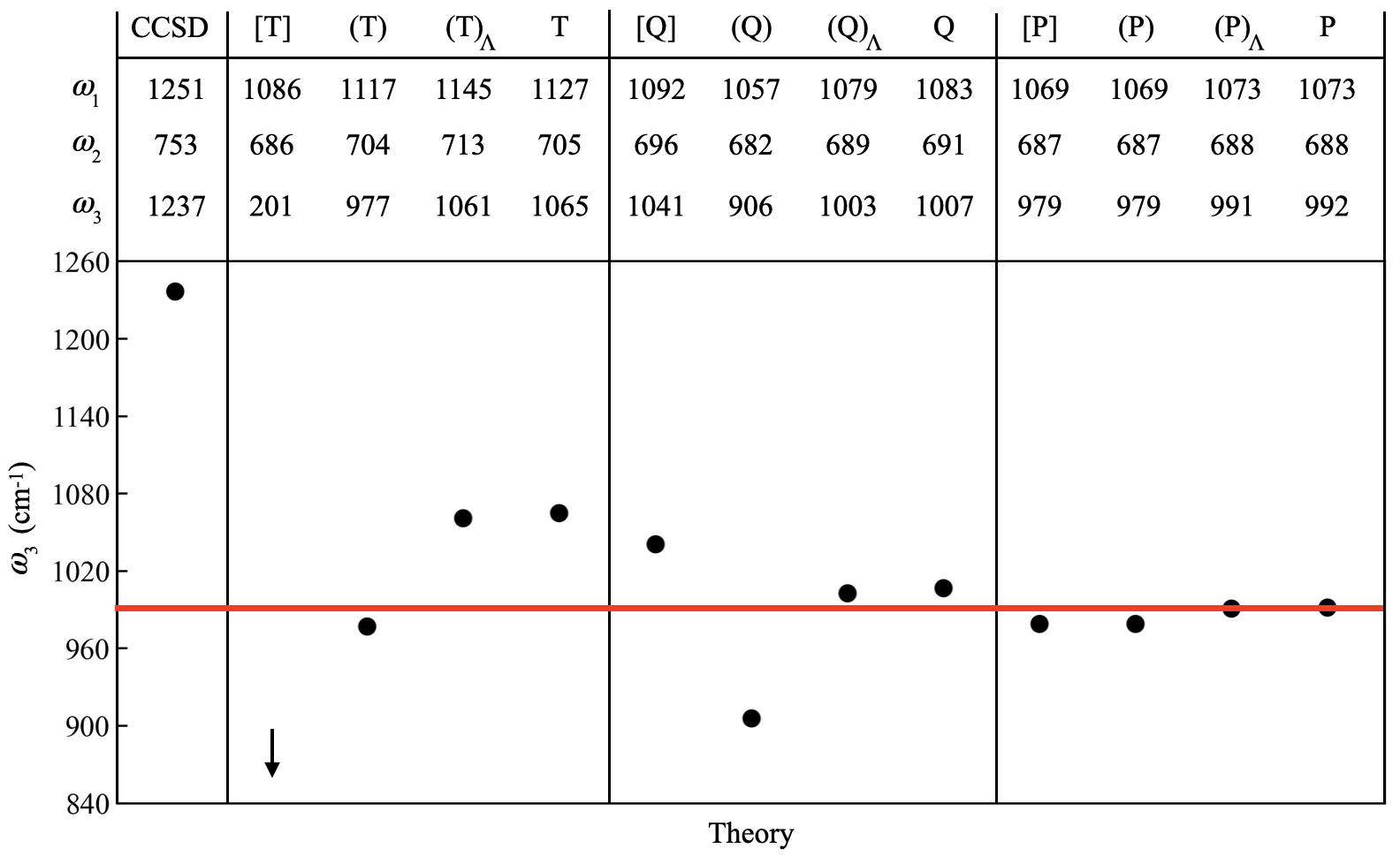}
\end{figure}

It is clear from Figure \ref{fig:megan} and Table \ref{tab:ozone} that the asymmetric (lambda-based) methods provide harmonic frequencies in better agreement with the corresponding fully iterative methods (such as CCSDT) than the more pragmatic symmetric approaches (like CCSD(T)). In this context, it is notable that CCSDT(Q)$_{\Lambda}$ agrees extremely well with the much more expensive CCSDTQ approach, while CCSDT(Q) fares slightly worse than CCSD(T). The latter's popularity can be viewed in this context as benefiting from an overestimation of triples effects, which partially compensates for the higher excitations that are neglected by these methods. 
This error compensation is well known from computational thermochemistry (e.g., Refs.\cite{heat1,heat2,heat3,heat4,jmlm200,jmlm205,jmlm235,jmlm273}), but is known there to break down beyond a certain level of static correlation. For instance, in Table S2 in Ref.\cite{jmlm273}, one finds the following post-CCSD(T) contributions to the ozone total atomization energy: higher-order triples CCSDT -- CCSD(T) =-1.34 kcal/mol, connected quadruples $T_4$=+3.81 kcal/mol, and connected quintuples $T_5$=+0.41 kcal/mol. By way of contrast, for a few representative molecules with less severe static correlation, error cancellation is seen to be largely preserved: for CO$_2$, $T_3$-(T)=-1.03, $T_4$=+1.04, and $T_5$=+0.05 kcal/mol, and for H$_2$O$_2$: -0.55, +0.72, and +0.03 kcal/mol, respectively.
From the point of view of formal theory, it is gratifying that the derivable lambda-based corrections \cite{lambdastanton1,lambdastanton2,lambdabartlett1,lambdabartlett2} outperform the less theoretically justified symmetric methods (such as CCSD(T)). Nonetheless, it can be seen in Figure \ref{fig:megan} and Table \ref{tab:ozone} that CCSD(T) and CCSDT(Q)$_{\Lambda}$ offer two ``sweet spots" within the coupled cluster series that accurately approximate the harmonic frequencies calculated by CCSDTQ5, a trend that was also observed for the diatomic test cases in the prior section. The reader should note that 
the T$_{5}$ contribution to the asymmetric-stretching harmonic frequency is still 16 cm$^{-1}$, which might lead some reader to question whether it is truly a reliable estimate of full CI.

In an attempt at reaching the latter independently through a multireference approach, we carried out full valence CASSCF-AQCC calculations using MOLPRO 2022.3\cite{Molpro2020}. As seen in Table \ref{tab:ozone}, the corresponding harmonic frequencies agree to 1 cm$^{-1}$ or better with CCSDTQ(5)$_{\Lambda}$, and the geometry to the third decimal place (in {\AA}ngstr\"om). 

Moreover, it can be seen in Table~\ref{tab:ozone} that as the basis set is increased all the way to cc-pV5Z, the full valence CASSCF-AQCC calculations closely approach experiment. 

Turning back to the cc-pVDZ basis set: replacing AQCC in the internally contracted multireference calculations by its direct ancestor ACPF (averaged coupled pair functional\cite{Gdanitz1988}) slightly degrades agreement with CCSDTQ(5)$_\Lambda$ (to 2 cm$^{-1}$), but represents no fundamental departure. Replacing AQCC by MRCI (multireference configuration interaction) without a size-consistency correction increases differences to +9 cm$^{-1}$; inclusion of a Davidson-type size-consistency correction\cite{Siegbahn1983,Davidson1974} reduces the discrepancy between CCSDTQ5(6) and CASSCF(18,12)-icMRCI+Q to 1 cm$^{-1}$ or less for the frequencies, while the bond distance agrees to four significant figures. We also note in passing that the CAS(2,2)-MRCCSDT (multireference coupled cluster theory) results of Hanauer and K\"ohn are remarkably close to fully iterative CCSDTQ.

For the cc-pVTZ basis set, we were able to obtain at most CCSDT(Q)$_\Lambda$ frequencies, as well as those with lower-level approaches. In fact, with this basis set, the superiority of CCSDT(Q)$_\Lambda$ over CCSDT(Q) is even more obvious (Table~\ref{tab:ozone}). Owing to the same error compensation as for cc-pVDZ, CCSD(T) actually outperforms CCSDT(Q).

We were alas unable to bring CCSDT(Q)$_\Lambda$/cc-pVQZ calculations to completion, but Table~\ref{tab:ozone} reveals that the deviations from CAS(18,12)-icAQCC of CCSD(T), CCSDT, and CCSDT[Q] are remarkably similar to those for the cc-pVTZ basis set, indicating that the latter have stabilized with respect to the basis set.

How well does full-valence CASSCF-AQCC perform for the diatomic test species of subsection~\ref{sec:diatomics}? We checked this for the cc-pVDZ basis set and found an RMSD with CCSDTQ5(6) of 3.2 cm$^{-1}$. This does include some cases such as F$_{2}$, ClF, and Cl$_{2}$, where a full-valence active space is inadequate as it only contains a single virtual orbital for these species; upon their removal or recalculation with some added virtual orbitals, this can be reduced somewhat further. Suffice it to say that the gap is already small enough that we have further evidence for our CCSDTQ(5)$_{\Lambda}$ 
ozone frequencies being in error by \textit{at most} a few wavenumbers. Thus, we recommend the CCSDTQ(5)$_{\Lambda}$ values to serve as a target of comparison for future benchmark calculations. 

In fact, as the revision of this paper was being prepared for submission, fully iterative CCSDTQ5/cc-pVDZ calculations finally achieved convergence to the required precision after months of wall clock time. As shown in Table~\ref{tab:ozone} at the bottom of the coupled cluster cc-pVDZ block, the differences with the CCSDTQ(5)$_\Lambda$ frequencies are just +0.3, +0.1, and +1.6 cm$^{-1}$, and in the geometry just -0.00003 {\AA} and 0.004 degree. This further drives home the point that CCSDTQ(5)$_\Lambda$ is the next `Pauling point' past CCSD(T) and CCSDT(Q)$_\Lambda$.

\section{Conclusions}

Our major conclusion for this dataset of diatomic spectroscopic constants is  that within the coupled cluster series, and taking computational cost into account, CCSD(T) and CCSDT(Q)$_\Lambda$ emerge as two `sweet spots' or Pauling points. In situations where static correlation is not too strong, CCSDT(Q) may be an acceptable alternative, with about one-half the memory demands and 2-3 shorter run times (owing to the elimination of the `left eigenvector').

CCSD(T) being widely regarded as `the gold standard of quantum chemistry' (a nickname first coined by Thom H. Dunning), perhaps CCSDT(Q)$_\Lambda$ may be considered the `platinum standard', and CCSDTQ(5)$_\Lambda$ a `diamond standard'. 

Finally, a composite of all-electron CCSD(T)-X2C/ACV5Z-X2C + [CCSDT(Q)$_\Lambda$ -- CCSD(T)]/cc-pVTZ would appear to be a felicitous combination for computational vibrational spectroscopy, and a further cost reduction by substitution cc-pVDZ for cc-pVTZ does not greatly impair accuracy: RMSD($\omega_e$)= increases from 2 to 2.5 cm$^{-1}$. These combinations may prove especially useful for polyatomics.

\section{ACKNOWLEDGMENTS}
Research at Weizmann was supported by the Israel Science Foundation (grant 1969/20), by the Minerva Foundation (grant 2020/05), and 
by a research grant from the Artificial Intelligence for Smart Materials Research Fund, in Memory of Dr. Uriel Arnon.
Research in Florida was supported by  the US Department of Energy (DOE), Office of Science, Office of Basic Energy Sciences (BES) under contract no. DE-FG02-07ER15884.  ES thanks the Feinberg Graduate School (Weizmann Institute of Science) for a doctoral fellowship and the Onassis Foundation (Scholarship ID: FZP 052-2/2021-2022). The visit of Maciej Spiegel to Weizmann was supported by the Erasmus student exchange program of the European Union. \\

This paper is dedicated to the memory of Dr. Timothy J. Lee (1960-2022).

\section*{Supporting information and data availability statement}

A Microsoft Excel workbook with results from the Dunham analysis is provided as Supporting Information. Additional data can be obtained from the authors upon reasonable request.

\bibliographystyle{aipnum-cp}
\bibliography{lambdafreq}
\end{document}